\documentclass[a4paper,twocolumn,english]{article}
\usepackage[T1]{fontenc}
\usepackage[latin9]{inputenc}
\usepackage{xcolor}
\usepackage{pdfcolmk}
\usepackage{babel}
\usepackage{bm}
\usepackage{amsmath}
\usepackage{graphicx}
\PassOptionsToPackage{normalem}{ulem}
\usepackage{ulem}
\usepackage[unicode=true]
 {hyperref}

\makeatletter

\pdfpageheight\paperheight
\pdfpagewidth\paperwidth

\providecolor{lyxadded}{rgb}{0,0,1}
\providecolor{lyxdeleted}{rgb}{1,0,0}
 \usepackage{amsmath}
 \usepackage{amssymb}
 \usepackage{bm}
\usepackage{graphics}
\usepackage{amssymb}
\usepackage{booktabs}

\makeatother

\begin{document}

\title{Efficient method of finding scaling exponents from\\
finite-size Monte-Carlo simulations%
\thanks{The final publication is available at \protect\href{http://www.epj.org}{www.epj.org}.%
}}

\author{I. Mandre and J. Kalda}
\maketitle
\begin{abstract}
Monte-Carlo simulations are routinely used for estimating the scaling
exponents of complex systems. However, due to finite-size effects,
determining the exponent values is often difficult and not reliable.
Here we present a novel technique of dealing with the problem of finite-size
scaling. This new method allows not only to decrease the uncertainties
of the scaling exponents, but makes it also possible to determine
the exponents of the asymptotic corrections to the scaling laws. The
efficiency of the technique is demonstrated by finding the scaling
exponent of uncorrelated percolation cluster hulls. 
\end{abstract}

\section{\label{sec:intro}Introduction}

Determining the scaling exponents from the finite-size simulation
data is a very common task in the physics of complex systems. In particular,
this technique is widely used in the context of phase transitions,
surface roughening, turbulence, granular media, etc, c.f. reviews
\cite{Barber,Binder,Privman}. Typically, such finite-size Monte-Carlo
studies involve extrapolation of the simulation data towards infinity.
Unless there is some theoretical understanding about the functional
form of the finite-size corrections to the asymptotic scaling laws
of the particular system, such an extrapolation carries a risk of
underestimating the uncertainties. In some cases, it may be helpful
to increase the computation time and system size, and optimize the
simulation scheme (c.f. \cite{JKMC}). However, this is not always
feasible, because the convergence to the asymptotic scaling law may
be very slow, c.f. \cite{JK08EPL}. Additional difficulties arise,
when one needs to determine the exponents of the finite-size correction
terms (c.f.~\cite{Bischof}), or when the asymptotic power law includes
a logarithmic pre-factor.

In what follows, we describe a novel technique for determining scaling
exponents from the finite-size simulation data. First, we describe
in which form the scaling law is expected to hold, and review the
traditional method. Then, we introduce the basic idea which allows
us to improve qualitatively the precision of the finite-size Monte-Carlo
studies, the idea of studying simultaneously multiple physical quantities
that asymptotically scale with the same exponent, but have different
finite-size correction terms. After that, we describe the novel method
to analyze Monte-Carlo simulation data for extracting the scaling
exponents and the finite-size correction terms. Finally, we provide
an example application of the technique and find the scaling exponent
of the uncorrelated percolation cluster hulls. A comparison is offered
with the naive application of fitting to the asymptotic scaling law
without considering the finite-size correction terms.

\section{\label{sec:old-way}The asymptotic scaling law}

Let us consider a system (possibly idealized, modeling a real one),
which is characterized by its size $x$, assuming that the smallest
possible value of $x$ plays the role of the unit length.

Further, suppose that the mathematical expectation of a certain physical
quantity scales as
\begin{equation}
\left\langle L\left(x\right)\right\rangle \propto x^{\alpha},\qquad x\gg1;\label{eq:L-scaling-law}
\end{equation}
here, the angular braces denote averaging over the full ensemble of
the model systems. The Monte-Carlo simulations can be used to estimate
the values of the mathematical expectation \eqref{eq:L-scaling-law}
for several system sizes $x_{1}<x_{2}<\ldots<x_{n}$, denoted as
\begin{equation}
\mathcal{L}_{i}\equiv\overline{L\left(x_{i}\right)},\qquad i=1\ldots n,
\end{equation}
and the variances of them as $\sigma_{i}^{2}$; the bar over a symbol
denotes averaging over a set of Monte-Carlo simulations. Then, a least-square
fit can be used to obtain the scaling exponent $\alpha$, c.f.~\cite{Binder}.
However, it is often difficult to estimate the uncertainty of the
obtained result, because the magnitude of the finite-size corrections
$\Delta$ within
\begin{equation}
\left\langle L\left(x\right)\right\rangle =Ax^{\alpha}+\Delta\left(x\right)\label{eq:simple-asymptotic-law}
\end{equation}
is unknown. Of course, one can plot $\ln{\cal L}_{i}$ versus $\ln x_{i}$
and determine such a crossover point $i=k$ that for $i\ge k$, the
data points lay within their statistical uncertainties on a straight
line. Then, only the data points with $i\ge k$ will be used for finding
the exponent $\alpha$. However, one can easily underestimate the
adequate value of $k$, because the statistical fluctuations just
happen to compensate the finite-size corrections $\Delta$. On the
other hand, taking excessively large values of $k$ would inflate
the variance of the outcome. Finally, in some cases, the decay rate
of the corrections $\Delta$ can be very slow, so that the method
outlined above will fail at the first step --- there is no linear
range of the graph. 

To resolve these problems we are going to make a series of assumptions.
Later we will see that the method we develop here also validates these
assumptions as it is applied and so the assumptions don't have to
be tested externally.

\textbf{First}, we assume a more complex scaling law for the mathematical
expectation of the physical quantity $L$, in the form
\begin{equation}
\left\langle L\left(x\right)\right\rangle =\sum_{k=1}^{\infty}A_{k}x^{\alpha_{k}},
\end{equation}
assuming that the most significant (in the sense of contributing to
the $\mathcal{L}_{i}$) members of the sum come first. The greatest
of the exponents $\alpha_{k}$ is the $\alpha$ we are looking for.
We separate $m$ first members and rewrite the sum as
\begin{equation}
\left\langle L\left(x\right)\right\rangle =\sum_{k=1}^{m}A_{k}x^{\alpha_{k}}+\Delta\left(x\right).
\end{equation}
This form for the finite-size correction terms has been used previously,
c.f. \cite{Ziff2011}.

\textbf{Second}, we assume that the contribution of $\Delta$ to $\mathcal{L}_{i}$
is smaller than their statistical fluctuation.

Now we can apply the least-squares fit to search for the $2m$ parameters,
$A_{k}$ and $\alpha_{k}$, $k=1\ldots m$. However, there are a few
problems. Unless we have some underlying idea about the parameters,
the least-squares search is complicated --- $m$ of the parameters
are non-linear and the search space is huge with many local minima.
We need at least $n\geq2m+1$ data points, all at different system
sizes --- increasing computational complexity. Also, we can't be sure
the assumptions we have made so far are actually correct (aside from
the chi-square test that is designed to test data probability rather
than the model).

\section{\label{sec:different-physical-quantities}Different physical quantities}

Our method is designed to resolve these problems; it will work, if
the following third\textbf{ }condition is satisfied.

\textbf{Third}, we assume that it is possible to find more than one
physical quantity with similar scaling behavior. So, we assume that
instead of having just one quantity, we can define $m$ distinct (linearly
independent in the finite scale) quantities, the mathematical expectations
$\left\langle L_{j}\right\rangle $ ($j=1\ldots m$) of which asymptotically
scale using the same exponent $\alpha$, but also have the same exponents
$\alpha_{k}$ ($k=1\ldots m$, so we have the same number of exponents
as physical quantities) for the finite-size correction terms:

\begin{equation}
\left\langle L_{j}\left(x\right)\right\rangle =\sum_{k=1}^{m}A_{jk}x^{\alpha_{k}}+\Delta_{j}\left(x\right),\quad j=1\ldots m.\label{eq:math-exp-diff-phys-qualit}
\end{equation}
We denote $\mathcal{L}_{ij}\equiv\overline{L_{j}\left(x_{i}\right)}$
with corresponding covariances $\Sigma_{ikl}=\mathrm{Cov}\left(\mathcal{L}_{ik},\mathcal{L}_{il}\right)$;
these covariances can be easily calculated during the Monte-Carlo
simulations. For each system size we then have a covariance matrix
$\bm{\Sigma}_{i}=\left(\Sigma_{ikl}\right)_{kl},\; i=1\ldots n;$
with corresponding inverse matrices $\mathbf{W}_{i}=\bm{\Sigma}_{i}^{-1}=\left(w_{ikl}\right)_{kl}$.
A least-squares fit can now be done by minimizing
\begin{equation}
\sum_{i=1}^{n}\sum_{j,k=1}^{m}\left(\mathcal{L}_{ij}-\sum_{l=1}^{m}A_{jl}x_{i}^{\alpha_{l}}\right)w_{ijk}\left(\mathcal{L}_{ik}-\sum_{l=1}^{m}A_{kl}x_{i}^{\alpha_{l}}\right),\label{eq:big-trad-minim}
\end{equation}
which at minimum is of chi-square distribution with $nm-m^{2}-m$
degrees of freedom. We have reduced the necessary calculation complexity
as we now only need $n\geq m+2$ different system sizes. Further,
the distinct physical quantities that scale using the same exponents
can be calculated from the same system instance within the Monte-Carlo
simulations.

The minimization problem is still non-linear in $m$ parameters and
now with total of $m^{2}+m$ parameters. We found it yields well to
the Levenberg\textendash{}Marquardt algorithm, given proper initial
values. However, with inadequate initial values, it can still lead
to inconsistent results and local minima.

It is trivial that more data should yield a better result. The third
assumption shows how to get this data and how it is done at no extra
computational cost. Next we look into how to consistently apply this
``free'' data to yield better results.

\section{\label{sec:new-method}Description of the method}

To simplify the problem we rewrite eq.~\eqref{eq:math-exp-diff-phys-qualit}
in matrix form, with $\mathbf{L}=\left(\left\langle L_{j}\left(x\right)\right\rangle \right),\ \mathbf{A}=\left(A_{jk}\right),\ \mathbf{X}=\left(x^{\alpha_{k}}\right),\ \bm{\Delta}=\left(\Delta_{j}\left(x\right)\right)$,
and derive
\begin{equation}
\begin{array}{cclcc}
\mathbf{L} & = & \mathbf{A}\mathbf{X}+\bm{\Delta},\\
\mathbf{X} & = & \mathbf{A}^{-1}\mathbf{L}-\mathbf{A}^{-1}\bm{\Delta} & = & \mathbf{B}\mathbf{L}+\bm{\delta},
\end{array}
\end{equation}
where $\mathbf{B}=\mathbf{A}^{-1}$ and $\bm{\delta}=-\mathbf{A}^{-1}\bm{\Delta}$.
A single row from this equation is
\begin{equation}
x^{\alpha_{k}}=\sum_{j=1}^{m}B_{kj}\left\langle L_{j}\left(x\right)\right\rangle +\delta_{k},\quad k=1\ldots m.
\end{equation}
We remark here that as $\Delta_{j}$ are small, so are the $\delta_{k}$.

We now attempt to find the parameters $B_{kj}$ by treating this as
a least-squares fitting problem. For this, we construct a function
\begin{equation}
S\left(d\right)=\sum_{i=1}^{n}\frac{\left(x_{i}^{d}-\sum_{j=1}^{m}C_{j}\mathcal{L}_{ij}\right)^{2}}{s_{i}^{2}}.\label{eq:method-S-eqation}
\end{equation}
The weighting factor $s^{2}$ is simply the variance of the expression
within the parentheses:
\begin{equation}
s_{i}^{2}=\mathrm{Var}\left(x_{i}^{d}-\sum_{j=1}^{m}C_{j}\mathcal{L}_{ij}\right)=\sum_{kl=1}^{m}C_{k}C_{l}\Sigma_{ikl}.
\end{equation}

We minimize the function $S\left(d\right)$ in relation to the parameters
$C_{1},\ldots,C_{m}$. Aside from the weighting factor $s^{2}$, that
depends on the values $C_{k}$, this is a simple linear-least-squares
problem. We found that by initially setting $C_{k}$ to $1$ and iteratively
running the linear-least-squares algorithm, then near the minima of
$S\left(d\right)$ the function value converges in three or four iterations.

Considering the assumptions made, it is clear that near $d=\alpha_{k}$
the function $S\left(d\right)$ should have a minimum. Conversely,
if the function $S\left(d\right)$ has exactly $m$ clear minima,
our assumptions about the scaling law must be correct and values of
$\alpha_{k}$ are exactly where $S\left(d\right)$ has minima. Hence,
we have found a way to extract the values $\alpha_{k}$ from the function
$S\left(d\right)$.

For statistical testing, the vectors $\left(\mathcal{L}_{i1},\ldots,\mathcal{L}_{im}\right),\ i=1\ldots n$
must be of multivariate normal distribution. Satisfying this, at minima
the function $S\left(d\right)$ is of chi-square distribution with
$n-m-1$ degrees of freedom. Consequently, just as with \eqref{eq:big-trad-minim},
we must have $n\geq m+2$. To accept the exponents $\alpha_{k}$ as
significant, a chi-square test must be performed: at minima the function
$S\left(d\right)$ has to satisfy the relation
\begin{equation}
S\left(\alpha_{k}\right)\leq\chi_{n-m-1}^{2}\left(p\right),
\end{equation}
where $\chi_{\mathrm{dof.}}^{2}\left(p\right)$ is the quantile at
$p$ of the chi-square distribution with $n-m-1$ degrees of freedom
($\mathrm{dof.}$).

Aside from the exponents $\alpha_{k}$, we can also find their uncertainties
$\Delta\alpha_{k}$ from
\begin{equation}
S\left(\alpha_{k}\pm\Delta\alpha_{k}\right)=S\left(\alpha_{k}\right)+\chi_{1}^{2}\left(p\right),\label{eq:method-stderr-relation}
\end{equation}
here we are making use of the constant chi-square boundary as the
confidence limit --- $\Delta\alpha_{k}$ is determined by the width
of the dip at the minimum of $S\left(d\right)$, at level $S\left(\alpha_{k}\right)+\chi_{i}^{2}\left(p\right)$.

In case we are uncertain about the results, we can always revert back
to \eqref{eq:big-trad-minim}. We found that when doing so, the parameters
derived using the above described novel method perform flawlessly
as initial values for this non-linear minimization problem and results
yielded by the classical but complex \eqref{eq:big-trad-minim} are
the same.

Compared to \eqref{eq:big-trad-minim}, where we have a nonlinear
multidimensional minimization problem, the novel method contains a
linear one-dimensional search. This gives us consistent results as
we don't have to deal with local minima. Furthermore, each of the
correction exponents is statistically tested separately, instead of
one big sum in \eqref{eq:big-trad-minim} -- we have found that this
excludes invalid results that would otherwise pass.

\section{\label{sec:example}Example application}

\begin{figure}
\noindent \begin{centering}
\includegraphics[width=0.99\columnwidth]{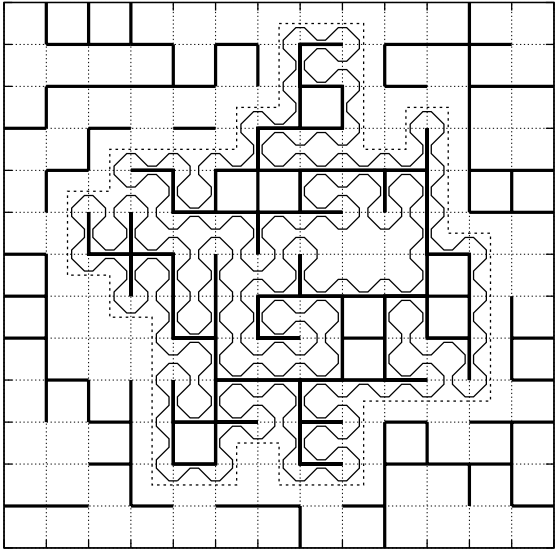}
\par\end{centering}

\caption{\label{fig:bond-percolation}Square bond percolation lattice. Bonds
(bold solid line segments) are randomly placed into the lattice. Clusters
are formed by bonds that are connected to each other. The largest
cluster in the center is illustrated with its hull (the zig-zag line)
and the unscreened perimeter (the dotted line).}
\end{figure}

As an example of the techniques described, we calculate the scaling
exponent of the hull of the uncorrelated percolation cluster. The
percolation problem deals with the structures that form by randomly
placing elementary geometrical objects (spheres, sticks, sites, bonds,
etc.) either freely into continuum, or into a fixed lattice (fig.~\ref{fig:bond-percolation}).
Two objects are said to communicate, if their distance is less than
some given $\lambda_{0}$, and communicating objects form bigger structures
called clusters. Percolation theory studies the formation of clusters
and their properties. The more interesting aspect is when and how
does an infinite cluster form. This depends on the lattice site occupation
probability. The minimum site occupation probability when an infinite
cluster appears is called the percolation threshold. Near this probability,
the percolation model displays critical behavior and long-range correlations.
For the square bond percolation model we use here, this critical probability
is $p=0.5$.

Percolation theory is used to study and model a wide variety of phenomena,
for example fluid flow in a porous medium \cite{Schmittbuhl2000},
thermal phase transitions and critical behavior in magnetism with
dilute Ising models \cite{Harris1974}. 

Several structures can be identified in conjunction with a percolation
cluster. For example, the cluster itself, the hull and the unscreened
perimeter (fig.~\ref{fig:bond-percolation}). Aside from these, many
others are known such as the oceanic coastline \cite{JK08EPL}, the
backbone or the chemical (shortest) distance. Near the percolation
threshold, all of these structures are fractals and can be characterized
by scaling exponents.

\begin{figure}
\noindent \begin{centering}
\includegraphics[width=0.99\columnwidth]{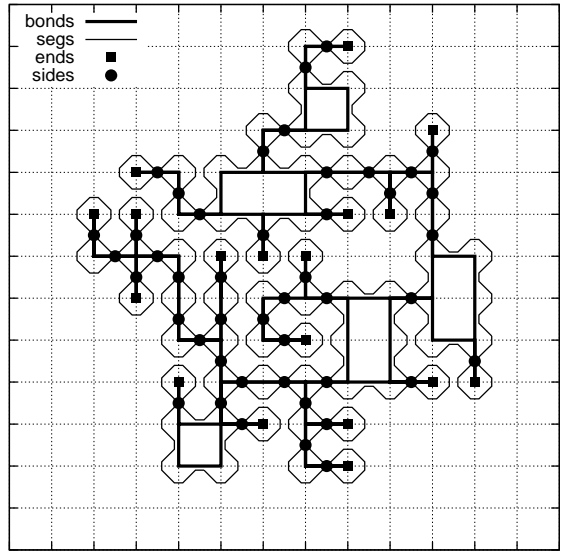}
\par\end{centering}

\caption{\label{fig:hull-properties}Some of the different physical quantities
that scale with the same exponent as the hull.}
\end{figure}

In this example, we concentrate on the scaling exponent of the hull
of uncorrelated percolation clusters. The exact value of this scaling
exponent is known, $d_{H}=1.75$ \cite{Saleur}; c.f.~\cite{MBI}.

First, we identify the different physical quantities (from here on,
the properties of the hull) that scale together with the hull. They
are (see fig.~\ref{fig:hull-properties})
\begin{itemize}
\item bonds -- the number of distinct bonds the hull touches,
\item segs -- the number of segments in the hull zig-zag,
\item ends -- the number of distinct bonds touched by the hull that have
no connections on one end,
\item sides -- the number of distinct bonds that are touched by the hull
from both sides,
\item lines -- the number of occurrences of four straight segments in the
hull,
\item corners -- the number of times bonds form corners in the hull,
\item ones -- the number of unset bonds by the hull that have exactly one
set bond connected to them,
\item twos -- the number of unset bonds by the hull that have exactly two
set bonds connected to them,
\item threes -- the number of unset bonds by the hull that have exactly
three set bonds connected to them.
\end{itemize}
It is possible to visualize how the scaling of these properties converges
towards the $d_{H}=7/4$. From \eqref{eq:L-scaling-law},
\begin{equation}
\mathcal{L}_{ij}\simeq C\cdot x_{i}^{\tilde{d_{j}}},\;\mathcal{L}_{\left(i+1\right)j}\simeq C\cdot x_{i+1}^{\tilde{d_{j}}},
\end{equation}
where $C$ is some constant. Dividing these two equations yields us
\begin{equation}
\frac{\mathcal{L}_{\left(i+1\right)j}}{\mathcal{L}_{ij}}\simeq\frac{x_{i+1}^{\tilde{d_{j}}}}{x_{i}^{\tilde{d}_{j}}}\Rightarrow\tilde{d_{j}}\simeq\ln\frac{\mathcal{L}_{\left(i+1\right)j}}{\mathcal{L}_{ij}}/\ln\frac{x_{i+1}}{x_{i}}.
\end{equation}
In simulations one often takes $x_{i+1}=2x_{i}$, and placing the
intermediate exponent at $\sqrt{x_{i+1}x_{i}}$, we get
\begin{equation}
\tilde{d}_{j}\left(\sqrt{x_{i}x_{i+1}}\right)=\ln_{2}\frac{\mathcal{L}_{\left(i+1\right)j}}{\mathcal{L}_{ij}},\qquad\left(x_{i+1}=2x_{i}\right).
\end{equation}

\begin{figure}
\noindent \begin{centering}
\includegraphics[width=0.99\columnwidth]{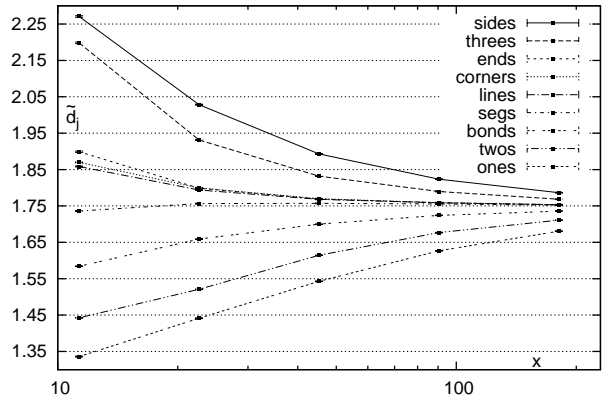}
\par\end{centering}

\caption{\label{fig:prop-convergence}Convergence of the scaling exponents
of the hull properties towards $d_{H}=1.75$.}
\end{figure}

The convergence of the nine studied properties towards the value $d_{H}=1.75$
can be seen in fig.~\ref{fig:prop-convergence}. The finite-size
effects are well pronounced for small system sizes. This data is practically
unusable for the simple model \eqref{eq:simple-asymptotic-law} ---
there is no linear range for the data values and any attempt will
fail at the chi-square test.

Some of the properties converge faster than others. Our method is
designed to work even with the very slowly converging properties.
Hence, to show its efficacy, out of the nine studied, we have selected
the five worst converging properties for what follows (sides, threes,
bonds, twos, ones).

\begin{figure}
\noindent \begin{centering}
\includegraphics[width=0.99\columnwidth]{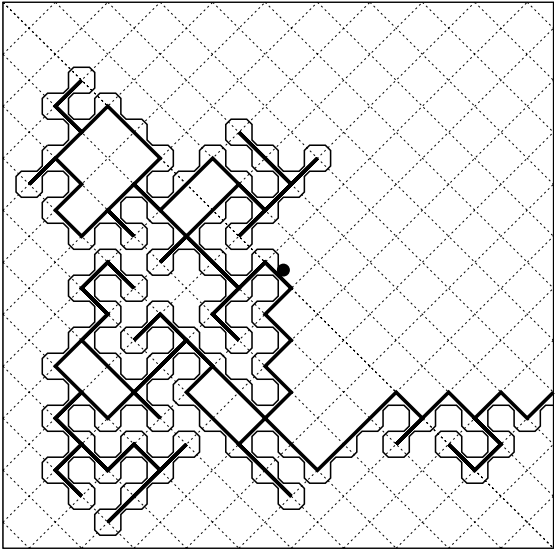}
\par\end{centering}

\caption{\label{fig:system-instance}Monte-Carlo simulation system instance
for scale length $x_{i}$. We start from the center (marked by a dot)
of an $x_{i}\times x_{i}$ box (for simplified bond coordinates we
use 45 degrees rotated lattice) and trace the hull until it reaches
an edge. Bond values are calculated dynamically on the way (from a
simple boolean random generator for the uncorrelated percolation).
We reject hulls that make a loop and so don't reach an edge. Various
hull properties are counted (for $\mathcal{L}_{ij}$) and their cross-multiplications
are calculated (for $\Sigma_{ijk}$). This is repeated for millions
of times for a single system size and the resulting data is aggregated.
Finally, $\mathcal{L}_{ij}$ and $\Sigma_{ijk}$ are calculated.}
\end{figure}

We run a Monte-Carlo simulation to gather data (the values $\mathcal{L}_{ij}$
and $\Sigma_{ijk}$ where $i=1\ldots n$ and $j,k=1\ldots m$; $n\geq m+2$).
This is done by tracing instances of hulls within the confines of
a system-sized box (fig.~\ref{fig:system-instance}). The system
sizes used were $8,\,16,\,\ldots,\,256$. At each system size $4.2\times10^{6}$
different hulls were generated and their properties counted.

\begin{figure}
\noindent \begin{centering}
\includegraphics[width=0.99\columnwidth]{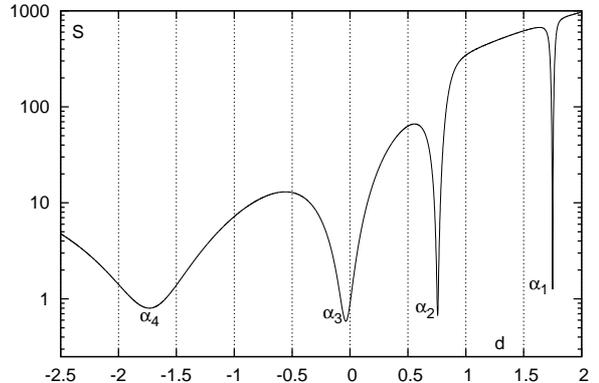}
\par\end{centering}

\caption{\label{fig:peaks}A sample uncorrelated percolation hull exponent
fitting function $S\left(d\right)$ using four different properties
($m=4$) of the percolation cluster hull (twos, segs, sides, ends).
The dips in the graph correspond to the exponents in \eqref{eq:math-exp-diff-phys-qualit}.
For this particular example, they are $\alpha_{1}=1.7494\pm0.0019,\,\alpha_{2}=0.756\pm0.018,\,\alpha_{3}=-0.04\pm0.16\mbox{ and }\alpha_{4}=-1.73\pm0.75$.}
\end{figure}

Once we have the data, we try out different variations of $m$ physical
quantities and find an instance of $S\left(d\right)$ that matches
our requirements (has $m$ clear minima that all satisfy the chi-square
test with $n-m-1$ degrees of freedom). One such combination (with
$m=4$) can be seen in fig.~\ref{fig:peaks}. The rightmost peak
is at the exponent $\alpha$ we are looking for and we can determine
its statistical uncertainty using relation \eqref{eq:method-stderr-relation}.

The number of exponents extractable is unknown, so different values
of $m$ must be tested. The chi-square test at the peaks may fail
if the statistical uncertainty in $\mathcal{L}_{ij}$ is comparable
to $\Delta_{j}\left(x_{i}\right)$ within \eqref{eq:math-exp-diff-phys-qualit}.
In such a case we must discard simulated data from the smallest system
and possibly run Monte-Carlo simulations for an additional larger
system. When discarding smaller systems, the constitution of the first
$m$ members in \eqref{eq:math-exp-diff-phys-qualit} may change ---
some members may only be significant for the smaller systems. When
that happens we may lose one or more of the minima and have to decrease
$m$. Parameter $m$ also determines the number of degrees of freedom
for the overall system (as we take $n=m+2$), hence while increasing
$m$ will decrease the contribution of the leftover finite-size correction
terms to the error (systematic error), it may at the same time slightly
increase the purely statistical uncertainty of the results.

\begin{table}
\noindent \begin{centering}
{
\small
\begin{tabular}{rlcccc}
  \toprule
    & Name & Smallest $\Delta d_H$ & Largest $\Delta d_H$ \\ % & Smallest error & Biggest error \\
  \midrule
  $1$ & LSQ 3 & $1.7299 \pm 0.0066$ & $1.653 \pm 0.031$ \\
  $2$ & LSQ 4 & $1.720 \pm 0.011$ & $1.619 \pm 0.044$ \\
  $3$ & MLSQ 2 & $1.7491 \pm 0.0011$ & $1.7488 \pm 0.0011$ \\
  $4$ & MLSQ 3 & $1.7492 \pm 0.0017$ & $1.7492 \pm 0.0017$ \\
  $5$ & MLSQ 4 & $1.7494 \pm 0.0019$ & $1.7492 \pm 0.0018$ \\
  \bottomrule
\end{tabular}
}
\par\end{centering}

\caption{\label{tab:comparing-methods}Results comparing fitting to the simple
model \eqref{eq:simple-asymptotic-law} versus the novel method ($\Delta d_{H}$
is the difference between the calculated and the known value). Only
first $6$ data points at $8,\,16,\,\ldots,\,256$ are used. LSQ N
- regular least squares fitting against model \eqref{eq:simple-asymptotic-law}
with one hull property and $N$ system sizes. MLSQ M - method described
in this paper, with $M$ different hull properties and $M+2$ system
sizes (as $M$ increases so does the system's degrees of freedom,
hence the uncertainty grows). Uncertainties are given with $0.95$
confidence. Note that none of the LSQ results passed the chi-square
test. The novel method offers consistent and accurate results.}
\end{table}

We can now compare the results from using the simple model (eq.~\eqref{eq:simple-asymptotic-law})
against the one one with $m$ different properties (eq.~\eqref{eq:math-exp-diff-phys-qualit}).
Results can be seen in table~\ref{tab:comparing-methods}. The method
offers correct results (within the confines of the statistical uncertainty),
high precision (small uncertainty) and consistent results {[}each
accepted $S\left(d\right)$, that is each combination of hull properties,
yields similar results{]}.

To be fair the gathered data is actually unusable for the simple model.
This is due to the finite-size correction terms. To make use of the
simple model \eqref{eq:simple-asymptotic-law} we would have to gather
data at much larger system sizes. To reach similar results (low statistical
error) to the novel method would demand vastly greater computational
costs.

Aside from the scaling exponent of the hull, we have also tested the
method to calculate the exponents of the unscreened perimeter $d_{U}=4/3$
and the cluster $d_{C}=91/48$ and obtained similar results to what
has been demonstrated above; the novel algorithm performed flawlessly
for all the cases. Finally, we have also studied the case of correlated
percolation, when the scaling exponents depend on the roughness (Hurst)
exponent $H$, so that $d_{H}=d_{H}\left(H\right)$. It is analytically
known that $d_{H}\left(0\right)=1.5$ \cite{Kondev1}; we have used
our method to recover this result with a high degree of precision
\cite{Mandre2011}.

In earlier studies \cite{Ziff2011,Aharony,Asikainen2003}, the correction
term exponents have been conjectured theoretically. When compared
to these studies, our results confirm the presence of the simple correction
terms (resulting from how we determine the diameter of a cluster and
also from constant offsets to the measurements of hull properties).
The inherent correction exponents described in those papers attributed
to percolation cluster scaling have not been found here. The most
likely explanation is that they were statistically insignificant.

\section{\label{sec:conclusion}Conclusion}

A novel and universal method of determining the scaling exponents
via finite-size Monte-Carlo simulations has been devised%
\footnote{An implementation can be found at https://code.google.com/p/perc2/,
see the 'calc' utility.%
}. The method can be applied, if it is possible to find $m\ge2$ distinct
quantities with equal asymptotic scaling exponents. The basic idea
is to exploit the equality of the exponents of finite-size correction
terms within the different physical quantities.

As an example, we have used the method to find the scaling exponents
of the uncorrelated percolation cluster hulls. Here the method offered
consistent results and increases the accuracy of the scaling exponent
estimates. The method has also been used previously in various contexts
in the field with good results, c.f. \cite{JK08EPL,Mandre2011}.

The method is particularly useful when the convergence to the asymptotic
scaling law is slow as it vastly reduces computational costs compared
to traditional methods. We can make use of small system sizes that
with traditional methods yield erroneous results or fail altogether.
Also, the method is extremely useful, if it is necessary to find the
exponents of the finite-size correction terms.

\section*{Acknowledgments}

This work was supported by Estonian Science Foundation grants No 7909,
Estonian Science Targeted Project No. SF0140077s08, and EU Regional
Development Fund Centre of Excellence TK124.

\bibliographystyle{plunsrt}
\bibliography{finsizescal}

\end{document}